\documentclass[amsmath,amssymb]{revtex4-2}
\usepackage[english] {babel}
\usepackage{amsfonts,graphicx,soul,natbib,mathrsfs,bm}

\begin{document}

\title{Turn-Key Constrained Parameter Space Exploration for Particle Accelerators Using Bayesian Active Learning}

\author{Ryan Roussel}%
\email{rroussel@uchicago.edu}
\author{Juan Pablo Gonzalez-Aguilera}%
\author{Young-Kee Kim}
\affiliation{Department of Physics, University of Chicago, Chicago, Illinois 60637, USA}%

\author{Eric Wisniewski}
\author{Wanming Liu}
\author{Philippe Piot}
\author{John Power}
\affiliation{Argonne Wakefield Accelerator, Argonne National Laboratory, Lemont, Illinois, 60439, USA}

\author{Adi Hanuka}%
\author{Auralee Edelen}%
\affiliation{SLAC National Laboratory, Menlo Park, California 94025, USA}%

\date{\today}

\begin{abstract}

Particle accelerators are invaluable discovery engines in the chemical, biological and physical sciences.
Characterization of the accelerated beam response to accelerator input parameters is often the first step when conducting accelerator-based experiments.
Currently used techniques for characterization, such as grid-like parameter sampling scans, become impractical when extended to higher dimensional input spaces, when complicated measurement constraints are present, or prior information is known about the beam response is scarce.
In this work, we describe an adaptation of the popular Bayesian optimization algorithm, which enables a turn-key exploration algorithm that replaces parameter scans and minimizes prior information needed about the measurement's behavior and associated measurement constraints.
We experimentally demonstrate that our algorithm autonomously conducts an adaptive, multi-parameter exploration of input parameter space, while navigating a highly constrained, single-shot beam phase-space measurement.
In addition to applications in accelerator-based scientific experiments, this algorithm addresses challenges shared by many scientific disciplines and is thus applicable to autonomously conducting experiments over a broad range of research topics.

\end{abstract}

\maketitle

\section{\label{sec:intro}Introduction}
Particle accelerators have enabled ground-breaking discoveries in the fields of chemistry \cite{colletier_novo_2016}, biology \cite{young_structure_2016} and physics \cite{jiang_origin_2016,singer_photoinduced_2016}.
They are also increasingly deployed for societal applications, such as in medical \cite{haberer_advances_2002} or industrial \cite{hamm_industrial_2012} fields.
During operation, accelerator parameters need to be tuned to produce beams with specific characteristics that match the needs for front-end applications.
Measuring these beam properties as a function of one or more input parameters using limited diagnostics and time-consuming measurements is a necessary part of operations, experimental planning, and tolerance determination, which comes at the expense of reducing accelerator availability for experimenters. 
These challenges are shared by many different scientific fields, which try to characterize complex, highly nonlinear and correlated systems, using difficult to execute scientific measurements and complicated diagnostics \cite{heath_introduction_1995,murray_application_2015}.

Due to the complex and time-consuming nature of accelerator diagnostics, characterization of beam response to input parameters is often limited to simple, uniformly spaced, grid-like parameter scans in one or two dimensions.
This limitation results from the poor scaling of grid-like scans to higher dimensional spaces, where the number of samples grows exponentially with the number of input parameters.
Furthermore, despite its simplicity at face value, it is often difficult to determine the ideal properties of parametric scans which will result in successful and efficient sampling.
A predefined grid spacing ultimately limits the ability to resolve fine features while oversampling slow variations of the measured parameters.
As a result, specifying the scan parameters $\textit{a priori}$ requires prior information about the measurement's functional dependence on each parameter.
This slows down frequent routine studies and makes characterization of novel measurements difficult to execute successfully.

The existence of tight constraints on which measurements are viable further complicates this process.
Upper and lower parameter limits are often determined by practical constraints of conducting measurements.
For example, transverse beam size measurements on diagnostic screens are limited by the screen size (available field of view), which in turn, imposes limits on the strength of upstream focusing magnet parameters.
Simulation studies or extra measurements are needed beforehand to determine these limits.
Even when simulations are available, they do not necessarily represent realistic machine behavior and measurements may become inaccurate due to time dependent changes in the accelerator, further complicating this problem.
Even if these limits can be determined for a single parameter, it becomes practically infeasible to efficiently determine limits in higher dimensional spaces, as they are often correlated with multiple parameters.
Limitations such as these are shared among many types of scientific experiments \cite{baltz_achievement_2017}.

Our goal is to produce an algorithm that replaces simple parametric scans for function characterization, while meeting potentially poorly understood requirements for practical measurements in a flexible manner.
The algorithm should be ``turn-key", requiring as little prior information and oversight as possible.
An ideal algorithm will adapt its sampling strategy to the observed functional behavior in order to increase sampling efficiency.
It should also determine the regions of input space which yield unsuccessful measurements and use this information to avoid invalid measurements in the future.
This algorithm reduces beamline tuning time during normal operations, while enabling efficient exploration of novel or poorly characterized systems.

We introduce and demonstrate an algorithm, coined ``Constrained Bayesian Exploration" (CBE), that solves this problem.
Starting from an initial valid observation, the algorithm sequentially samples a quasi-uniform grid of points in input space.
It adapts its sampling strategy to measured functional behavior with respect to each parameter and respects necessary constraints for successful measurements.
Finally, the algorithm is biased towards making small jumps in input space, which balances the trade-off between exploration and costs associated with changing input parameters.
While our focus here is the application of this method towards parametric accelerator exploration, this algorithm can be applied in any experimental scenario that requires parametric scans.

Our algorithm is an adaptation of Bayesian optimization (BO) \cite{shahriari_taking_2016, greenhill_bayesian_2020}, applied to maximizing information gain \cite{srinivas_gaussian_2010}, which is often referred to as active learning or uncertainty sampling \cite{duplyakin_active_2016, smith_less_2018, lookman_active_2019, settles_active_2009}. 
Bayesian optimization generally consists of two components.
The first component of BO is a probabilistic surrogate model that predicts the probability distribution of the function value $f$ as a function of the input parameter vector $\mathbf{x}$.
Gaussian Processes (GPs) \cite{rasmussen_gaussian_2006} are a popular model choice for Bayesian optimization.
The behavior of a GP model is determined by a kernel function, which describes the correlation between function values based on their location in input space. 
The kernel function itself can depend on a collection of hyperparameters, which most notably often includes a length scale parameter.
This parameter describes the effective ``smoothness" of the model, where small and large values correspond to rapidly- and slowly-varying functional behavior respectively.
Further, an independent length scale hyperparamter can be specified for each input parameter, in a process known as automatic relevance determination (ARD) \cite{neal_bayesian_2012}.
These hyperparameters are usually determined by maximizing the marginal log likelihood of the GP model \cite{rasmussen_gaussian_2006}, conditioned on experimental measurements which balances model accuracy and complexity.

The second component of BO is an acquisition function, which characterizes the value gained by observing a particular point in input space.
Bayesian optimization selects the next experimental sample by finding the point in input space that maximizes the acquisition function.
Several different acquisition functions that are commonly used for optimization are Probability of Improvement \cite{kushner_new_1964}, Expected Improvement \cite{mockus_application_1978} and Upper Confidence Bound (UCB) \cite{srinivas_gaussian_2010}. 
The UCB acquisition function is defined as
\begin{equation}
    \alpha(\mathbf{x}) = \mu(\mathbf{x}) + \sqrt{\beta} \sigma(\mathbf{x})
\end{equation}
where $\mu(\mathbf{x})$ is the predicted mean of the function value and $\sigma(\mathbf{x})$ is the predicted uncertainty, both determined by the GP model.
This acquisition function is of particular interest as it allows the user to specify the optimization parameter $\beta$, which represents the trade off between exploitation (sampling to take advantage of predicted extrema) and exploration (sampling to reduce prediction uncertainty).
An optimizer with a small value of $\beta$ prefers exploitation, while a large value of $\beta$ prioritizes exploration.

\begin{figure}
	\centering
	\includegraphics[width=0.5\linewidth]{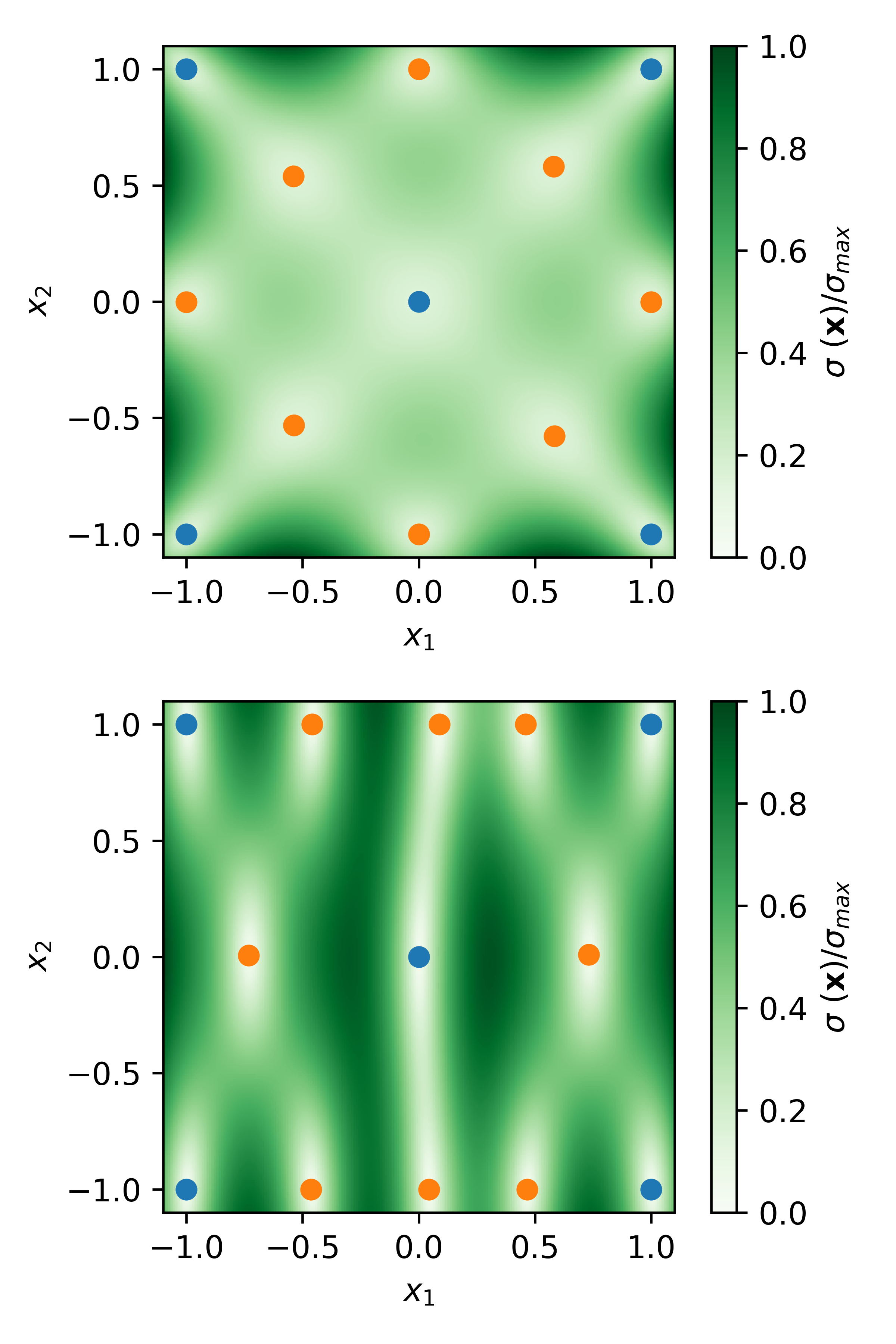}
	\caption{\label{fig:lengthscale} (Color) Plots showing Bayesian optimization sampling patterns depending on the kernel length scale using $\alpha(\mathbf{x}) = \sigma(\mathbf{x})$. Blue points are initial samples and orange points are determined via Bayesian optimization. 
	(Top) Length scales for both $[x_1,x_2]$ are set to 1. (Bottom) Length scales for variables $[x_1,x_2]$ are set to $[0.25,1]$.}
\end{figure}

It has been shown that maximum information gain of the GP model occurs when the acquisition function is only comprised of the uncertainty term $\alpha(\mathbf{x}) = \sigma(\mathbf{x})$ \cite{srinivas_gaussian_2010}, effectively choosing $\beta\rightarrow\infty$.
We note that the uncertainty predicted by the GP model is dependant only on the location of sampled points in input space (see Eqn.~\ref{eqn:uncert} in Methods for details).
Figure \ref{fig:lengthscale} shows the effect of kernel length scales on Bayesian optimization when we use this acquisition function.
If the GP kernel has a single length scale then the acquisition function is maximized at points which maximize the distance away from all previous observations.
The sampling algorithm will choose points that form a grid like pattern in input space, dependent on where initial samples were taken.
However, if the length scale is different along each input axis, the exploration algorithm will be biased towards sampling points along the axis associated with the shortest length scale.
This leads to an efficient sampling strategy as more samples are needed to resolve rapidly changing features of the function while fewer samples are used when the function changes slowly.
We use this behavior as a starting point for creating the CBE acquisition function.

We define the CBE acquisition function to be 
\begin{align}
    \alpha(\mathbf{x},\mathbf{x_0}) =& \sigma(\mathbf{x})\Psi(\mathbf{x},\mathbf{x_0})\prod_{i=1}^N P_i[g_i(\mathbf{x}) \geq h_i]\label{eqn:be_acq}\\
    \Psi(\mathbf{x},\mathbf{x_0}) =& \exp\Big(-\frac{1}{2}(\mathbf{x} - \mathbf{x_0})^T\mathbf{\Sigma}^{-1}(\mathbf{x} - \mathbf{x_0})\Big)
\end{align}
where the two additional terms have the following effects.

The term $\Psi(\mathbf{x},\mathbf{x}_0)$ in Eqn.~\ref{eqn:be_acq} represents a proximal biasing term where $\mathbf{x}_0$ is the most recently sampled point in input space.
This term is included for two reasons.
First, changing parameters in a real accelerator generally incurs a temporal cost, proportional to the change in the parameter.
Furthermore, changing accelerator parameters quickly can disrupt rapid feedback systems used to maintain supporting accelerator subsystems (such as those which maintain radio-frequency phase or power).
Second, we assume that the viable region is connected so prioritizing nearby points in input space increases the fraction of successful measurements vs. unsuccessful ones, as it is less likely to travel outside the valid region.
By including the proximal term, we negatively bias the acquisition function away from points that are not within a close proximity to the most recently observed point while still allowing exploration in cases where large jumps in input space result in highly valued observations.
The covariance matrix $\mathbf{\Sigma}$ specifies the length scale at which points are biased, where a smaller element in the precision matrix leads to a stronger biasing.

The last term, $\prod_{i=1}^N P_i[g_i(\mathbf{x}) \geq h_i]$ represents the multiplicative probability that $N$ operational constraints are satisfied, following \cite{gardner_bayesian_2014}.
This term weights the CBE acquisition function by the probability that a constraining function $g_i(\mathbf{x})$ is greater than a predetermined scalar value $h_i$.
As a result, this term will bias our acquisition function against sampling in regions of input space that are not likely to satisfy the constraints.
The probability is calculated based on GP model predictions of $\mu_i(\mathbf{x})$ and $\sigma_i(\mathbf{x})$ of the individual constraining functions $g_i(\mathbf{x})$, giving
\begin{equation}
    P_i[g_i(\mathbf{x})\geq h_i] = 1 - \Phi\Big(\frac{h_i - \mu_i(\mathbf{x})}{\sigma_i(\mathbf{x})}\Big)
\end{equation}
where $\Phi(x)$ is the Gaussian cumulative distribution function. 

\begin{figure*}
	\centering
	\includegraphics[width=1.0\linewidth]{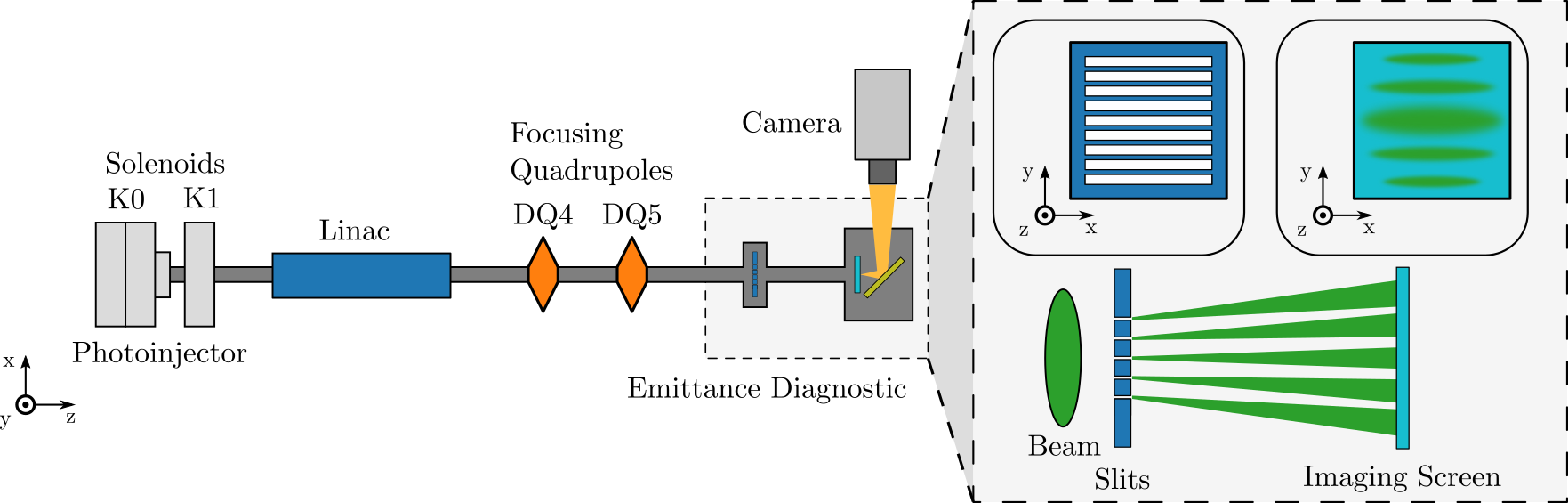}
	\caption{\label{fig:experiment_diagram} Cartoon depicting the emittance exploration experiment at the Argonne Wakefield Accelerator facility. Conceptual view of emittance diagnostic is shown on the right.
    }
\end{figure*}

\section{\label{sec:results} Results}
We conduced an experiment at the Argonne Wakefield Accelerator (AWA) to test how Bayesian exploration can be used to enable beamline characterization with a single shot emittance measurement.
The AWA beamline accelerates electrons to $\sim 42$~MeV using a photoinjector electron source combined with a normal-conducting linear accelerator~\cite{conde_research_2017}.
The transverse phase-space area or ``emittance" of beams produced by the accelerator is an important figure of merit that must be minimized to ensure ideal transport of the beam through the accelerator and meet specific experimental criteria. 
The emittance ultimately sets the beam brightness, a critical parameter in accelerator-based light sources \cite{huang_review_2007} and colliders \cite{grafstrom_luminosity_2015}.
The emittance is sensitive to several beamline parameters seen in Figure~\ref{fig:experiment_diagram}, including the magnetic field strength of a pair of solenoids surrounding the photoinjector (referred to here as the focusing solenoids and characterized by the scaling parameter K0) and a solenoid in between the photoinjector and the first accelerating cavity (referred to here as the matching solenoid and controlled via the scaling parameter K1).
Our goal is to explore the emittance response to these solenoids, as well as two quadrupole magnet located downstream of the linac. 
Too first order, these magnets control the transverse beam size without affecting the emittance \cite{hawkes_principles_1996}.
We use a single-shot multi-slit diagnostic \cite{zhang_emittance_1996}, also shown in Figure~\ref{fig:experiment_diagram}, to measure the beam emittance.
The principal challenge of conducting this measurement is that the beamline elements, while effecting the the beam emittance, also modify the beam size and divergence at the diagnostic.
Unfortunately, emittance measurements with this multi-split diagnostic are only possible over a narrow range of beam sizes and divergences.
Such a limited dynamical range is a common problem shared by many types of diagnostics across accelerator facilities.

\begin{figure*}
	\centering
	\includegraphics[width=1.0\linewidth]{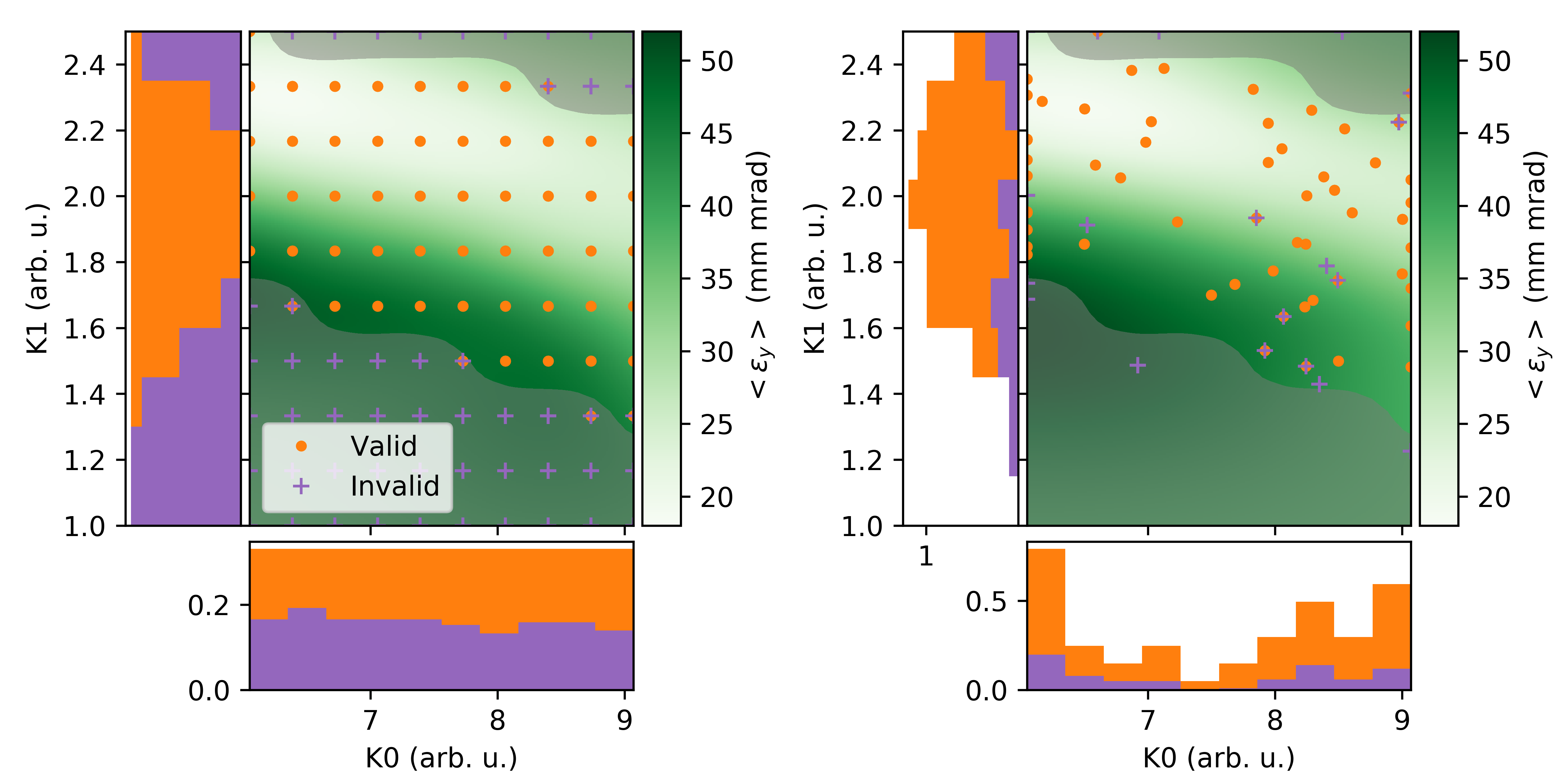}
	\caption{\label{fig:collected} (Color) Comparison between uniform grid sampling and Bayesian exploration. (Left) A 10 x 10 grid is sampled over an operational sub-domain identified by prior experimentation. Color mesh represents the posterior predicted mean of a Gaussian process trained on the valid measurements. Stacked histograms show sampling density of measurements along each axis. (Right) 2D projected samples from Bayesian exploration after 66 iterations with 4 parameters (K0, K1, DQ4, DQ5). In both plots, shading denotes invalid region in sub-domain with 95\% confidence from uniform grid sampling.
    }
\end{figure*}

As a baseline for comparison we conducted a 2-dimensional scan of the solenoid parameters (K0, K1).
A $10 \times 10$ point grid, shown in Figure \ref{fig:collected} was created with upper and lower bounds determined from prior experimentation.
For each set of parameters, five observations of the beam emittance were conducted while windowing the fluctuating bunch charge between 5.0 and 6.0 nC.
In some cases, points in input space result in both valid and invalid measurements due to accelerator noise.
We use these observations to train GP models of the vertical beam emittance $\varepsilon_y$ and the measurement validity constraining function $g(\mathbf{x})$, which is set to one for a successful measurement and zero otherwise.

From Figure \ref{fig:collected}, we observe that the invalid region of input space (denoted by shaded regions), defined as $P[g(\mathbf{x}) \geq 0.5]$, is roughly half the input space domain.
As a result, roughly half of our samples are wasted, as they provide no information on the beam emittance.
We observe that in the valid region the emittance is strongly dependant on K1 relative to K0.
If we normalize the input domain to the unit cube, the length scales of the GP kernel for each of these parameters is seen in Table \ref{table:lengthscales}.
The length scales inferred from the data are consistent with what is seen in Figure \ref{fig:collected} where the emittance changes slowly as a function of the focusing solenoid strength (K0) and quickly with respect to the matching solenoid strength (K1).

We then used CBE to explore a similar input space, varying the two solenoid magnet strengths $(K0,K1)$ and two quadrupole magnet strengths (DQ4, DQ5).
We initialized the algorithm with 2 initial valid measurements, randomly generated from within the valid sub-domain, determined from earlier experimentation.
The sigma matrix for the proximal term (given in normalized coordinate space) is set to $\mathbf{\Sigma} = 0.01 \mathbf{I}$ where $\mathbf{I}$ is the identity matrix.
As in the 2D uniform scan, we use the constraint inequality $g(\mathbf{x}) \geq 0.5$ where the constraint function $g(\mathbf{x}) = 1$ if the emittance measurement is valid and zero otherwise.

The results from this exploration, projected onto the 2D sub-space where DQ4 $=0$, DQ5 $=0$, appear in Figure \ref{fig:collected}.
We again plot the posterior predictive mean of the emittance GP model in this subspace, trained on the sampled data.
Estimated length scales from this model are shown in Table \ref{table:lengthscales}.
Similar to the uniform grid model, the length scale for K0 is significantly longer than the length scale for K1, implying that the function varies faster along K1, which is consistent with what is observed in Figure \ref{fig:collected}(b).
Further, we observe that the length scales for K0,K1 are similar when comparing the two models.
The difference between length scales for K1 is slightly larger, which could be explained by the larger average separation between points during grid sampling relative to CBE sampling.
Grid sampling places limits on the fitted length scale, where the spacing of uniform samples determines the fastest changes in the target function that can be observed.
Finally, the length scales for variables DQ4, DQ5 are larger than the normalized domain, implying that the emittance is weakly correlated with quadrupole-magnet strength, as expected by first order beam dynamics \cite{hawkes_principles_1996}.
We even observe that the length scale for DQ5 is significantly larger than the length scale associated with DQ4, which possibly results from the magnet's location in the beamline, closer to the emittance diagnostic, which reduces the impact of higher order effects on the emittance.  

The mean emittance of the Gaussian process model trained from exploration samples is also qualitatively similar to the model generated by grid sampling, despite the exploration algorithm using roughly 2/3 the number of samples.
A larger portion of samples taken by the exploration algorithm are valid compared to uniform grid sampling (77\% vs. 52\%) which provides more information about the emittance per sample on average.
Note that invalid samples which appear inside the valid region shown in Figure \ref{fig:collected} are invalid due to correlations with the un-plotted variables DQ4, DQ5.
Furthermore, from the projected histograms of K0 samples, we see that several regions of parameter space are avoided by the algorithm, due to the long length scale associated with K0 which reduces model uncertainty in between previous measurements. 
As a result, the algorithm skips over these regions that do not significantly reduce uncertainty if observed.
On the other hand, samples along the K1 axis, which has a much shorter length scale, are continuously distributed in the valid region.
This results in a better characterization the functional dependence on K1 since it shows rapidly changing behavior.

\begin{table}
    \caption{Trained hyperparameter length scales (normalized)}
    \label{table:lengthscales}
    \begin{ruledtabular}
        \begin {tabular}{l r r r}
            Parameter & Uniform grid sampling & Bayes. Exp.\\
    \colrule
    Focusing solenoid (K0) & 0.82 & 0.77\\
    Matching solenoid (K1) & 0.40 & 0.30\\
    DQ4 & N/A & 1.03\\
    DQ5 & N/A & 1.40
    \end {tabular}
    \end{ruledtabular}
\end{table} 

\begin{figure*}
	\centering
	\includegraphics[width=1.0\linewidth]{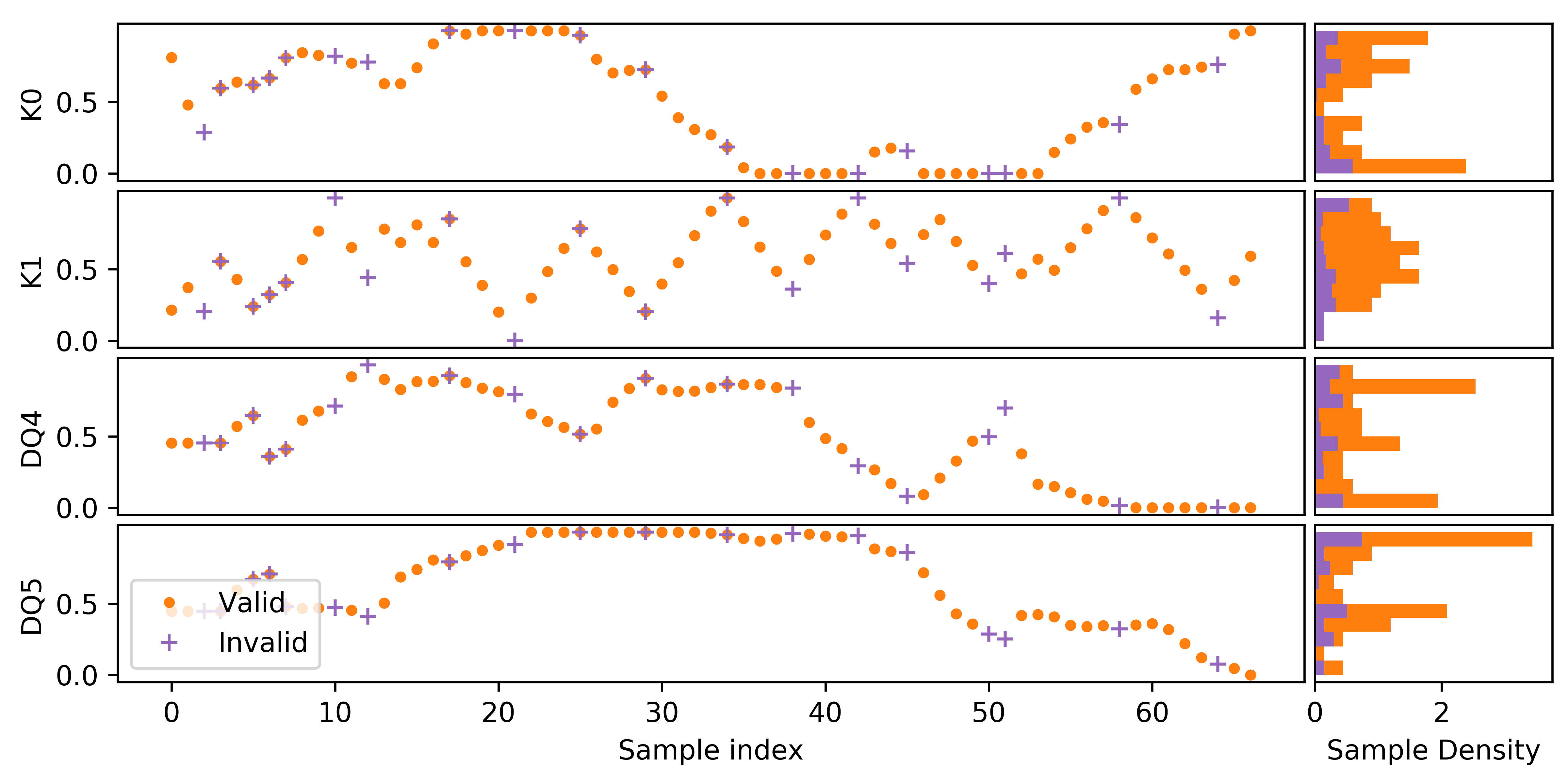}
	\caption{\label{fig:trace} (Color) Parameter values as a function of sample index during Bayesian exploration along with sample density histograms.}
\end{figure*}
The difference in sampling behavior is also observed in Figure \ref{fig:trace} where we plot the normalized trace of each parameter value during exploration.
As a result of the short length scale associated with K1, the algorithm effectively scans back and forth across the valid input region for this parameter while other parameters are slowly modified.
Furthermore, the maximum and minimum values of the scan are dynamically determined when the algorithm intersects with an invalid region.
This is important, as we clearly see that the valid region for K1 changes as a function of the other variables.
Finally, we observe that for variables with long length scales (K0, DQ4, DQ5), CBE will ignore intermediate regions that do not significantly reduce model uncertainty, reducing redundant measurements as it learns the associated length scale along these axes.

\section{\label{sec:conclusion} Discussion}
Here we have described and demonstrated a novel algorithm for autonomously and efficiently exploring input parameter spaces with limited prior information, while adapting to measurements that are potentially invalid or unsuccessful.
This is achieved without the need for prior information about the target function or measurement constraints.
We experimentally demonstrated our algorithm's ability to efficiently explore the functional dependence of the beam emittance on four accelerator parameters automatically, while navigating a difficult to execute measurement.
The Bayesian exploration algorithm was able to achieve similar model prediction accuracy as a grid based scan, with a smaller number of samples and two extra input parameters.
The advantages this algorithm confers over traditional methods of characterizing particle accelerator responses are substantial, and we expect it to have broad impact across accelerator-based science facilities.
While we demonstrated this algorithm's effectiveness in the context of particle accelerators, Bayesian exploration is a flexible, lightweight, turn-key algorithm that replaces the need for grid type parameter scans in any field.
In particular, the addition of a proximal biasing term in our sampling strategy has advantages over previously described active learning algorithms when exploring spatially dependent systems (robotics, surveying etc.) that encounter costs associated with travel through input space.

\section{Acknowledgements}
The authors would like to thank Sergei Nagaitsev for their discussions regarding the multi-slit diagnostic. This work was supported by the U.S. National Science Foundation under Award No. PHY-1549132, the Center for Bright Beams. The Argonne Wakefield Accelerator is supported by the U.S. Department of Energy, Office of High Energy Physics, under Contract No. DE-AC02-06CH11357.

\section{\label{sec:methods} Methods}
\subsection{Beam Emittance Diagnostic}
The geometric emittance of a beam can be determined experimentally using a multi-slit emittance diagnostic.
This diagnostic consists of a transverse mask with a number of horizontal slits that divides the beam into multiple ``beamlets".
A downstream transverse diagnostic screen is used to image the beamlets and the center of mass, size and integrated intensity of each beamlet is measured.
This information is used to calculate aspects of the beam envelope, and thus the geometric beam emittance.
To simplify our analysis we used a slight modification of the emittance formulas derived in \cite{zhang_emittance_1996}.
Instead of calculating the emittance using the correlated divergence, we used a calculation of the uncorrelated divergence to determine the emittance.
While this analysis prevents us from assessing the position-divergence correlation of the beam it still provides us with an accurate emittance measurement.

Our experiment used a laser etched stainless steel mask with 25 slits.
The slit pattern had a separation of 2 mm, a vertical slit width of 50~\textmu{m}, and a horizontal slit length of 40 mm.
The circular beam imaging screen, which had a diameter of 50~mm, was located 2.84 m downstream of the transverse mask.
The screen was imaged using an optical camera with a spatial resolution of 46~\textmu{m},  per pixel.

Emittance measurements were only considered valid if the measurement satisfied three conditions: we required that $(i)$  at least 5 beamlets were produced at the observation screen, $(ii)$ all the beamlets were contained within a predefined region of interest on the screen to prevent biasing due to potential clipping, and $(iii)$ the projection of each beamlet onto the vertical axis did not overlap with any other beamlet projections in order to properly measure the size of each beamlet.
If any of these requirements were not met, we assigned the constraining function at that location in input space a value of zero tagging the measurement as ``invalid".

\subsection{Gaussian Process Model Creation}
Non-parametric Gaussian process surrogate models are used to predict the value of a target function $f(\mathbf{x})$ using Bayesian statistics \cite{rasmussen_gaussian_2006}.
These models are specified by a covariance function, $k(\mathbf{x},\mathbf{x}';\bm{\phi})$ with hyperparameters $\bm{\phi}$ and a constant mean function $C$, such that we can write $f(\mathbf{x}) \sim \mathcal{GP}(C, k(\mathbf{x},\mathbf{x}'))$.
In an experimental setting, an observation $y$ is the target function corrupted by noise: $y = f(\mathbf{x}) + \epsilon$ where we assume that $\epsilon \sim \mathcal{N}(0,\sigma_{noise}^2)$.
Given $N$ previous measurements $\mathcal{D} = \{(\mathbf{x}_1,y_1),\dots,(\mathbf{x}_N,y_N)\}$ the predictive probability distribution of the function value $f = f(\mathbf{x})$ is given by 
\begin{equation}
p(f|\mathcal{D}, \mathbf{x}) = \mathcal{N}(\mu(\mathbf{x}),\sigma^2(\mathbf{x}))
\end{equation}
 where 
\begin{align}
    \mu(\mathbf{x}) =&\ \mathbf{k}^T[K + \sigma_{noise}^2 I]^{-1}(\mathbf{y} - C) + C\\
    \sigma(\mathbf{x}) =&\ k(\mathbf{x},\mathbf{x}) - \mathbf{k}^T[K + \sigma_{noise}^2 I]^{-1}\mathbf{k} \label{eqn:uncert}\\ 
    \mathbf{k} =&\ [k(\mathbf{x},\mathbf{x}_0),\dots,k(\mathbf{x},\mathbf{x}_{N})]^T\\
    K =&\ 
    \begin{bmatrix}
        k(\mathbf{x}_1,\mathbf{x}_1) & \cdots & k(\mathbf{x}_1,\mathbf{x}_N)\\
        \vdots & \ddots & \vdots\\
        k(\mathbf{x}_N,\mathbf{x}_1) & \cdots & k(\mathbf{x}_N,\mathbf{x}_N)
    \end{bmatrix}.
\end{align}
The model hyperparameters $\bm{\theta} = \{\phi, \sigma_{noise}, C\}$ are determined by maximizing the log marginal likelihood, $\bm{\theta}=\mathrm{argmax}_{\bm{\theta}}\log[p({D}; \bm{\theta})]$, which balances model accuracy and complexity when choosing hyperparameters.
A Mat\'{e}rn kernel ($\nu=3/2$) was used for the GP models in this paper.
During experimentation the hyperparameters were retrained after each observation.

\section{Author contributions}
R. R., J.P.G.A., A. E. wrote the manuscript; R. R., J.P.G.A. designed the experiment; R. R., W. L. wrote the software required to run the experiment; R. R., J.P.G.A, E. W., P. P. ran the experiment; J. P., P. P., A. H., A. E., Y. K. K. provided guidance and oversight, all authors contributed to the completion of the manuscript.

\section{Ethics declarations}
The authors declare no competing interests.

\bibliographystyle{unsrt}
\bibliography{BayesExp/BayesExp.bib}

\begin{thebibliography}{10}

\bibitem{colletier_novo_2016}
Jacques-Philippe Colletier, Michael~R. Sawaya, Mari Gingery, Jose~A. Rodriguez,
  Duilio Cascio, Aaron~S. Brewster, Tara Michels-Clark, Robert~H. Hice, Nicolas
  Coquelle, Sébastien Boutet, Garth~J. Williams, Marc Messerschmidt, Daniel~P.
  {DePonte}, Raymond~G. Sierra, Hartawan Laksmono, Jason~E. Koglin, Mark~S.
  Hunter, Hyun-Woo Park, Monarin Uervirojnangkoorn, Dennis~K. Bideshi, Axel~T.
  Brunger, Brian~A. Federici, Nicholas~K. Sauter, and David~S. Eisenberg.
\newblock De novo phasing with x-ray laser reveals mosquito larvicide {BinAB}
  structure.
\newblock 539(7627):43--47.
\newblock {ISBN}: 1476-4687.

\bibitem{young_structure_2016}
Iris D. et~al. Young.
\newblock Structure of photosystem {II} and substrate binding at room
  temperature.
\newblock 540(7633):453--457.
\newblock {ISBN}: 1476-4687.

\bibitem{jiang_origin_2016}
M.~P. Jiang, M.~Trigo, I.~Savić, S.~Fahy, É.~D. Murray, C.~Bray, J.~Clark,
  T.~Henighan, M.~Kozina, M.~Chollet, J.~M. Glownia, M.~C. Hoffmann, D.~Zhu,
  O.~Delaire, A.~F. May, B.~C. Sales, A.~M. Lindenberg, P.~Zalden, T.~Sato,
  R.~Merlin, and D.~A. Reis.
\newblock The origin of incipient ferroelectricity in lead telluride.
\newblock 7(1):12291.
\newblock {ISBN}: 2041-1723.

\bibitem{singer_photoinduced_2016}
A.~Singer, S.~K.~K. Patel, R.~Kukreja,
  V.~Uhlíıfmmode~\{{\textbackslash}textbackslash\}checkr\{{\textbackslash}textbackslash\}else
  ř\{{\textbackslash}textbackslash\}fi, J.~Wingert, S.~Festersen, D.~Zhu,
  J.~M. Glownia, H.~T. Lemke, S.~Nelson, M.~Kozina, K.~Rossnagel, M.~Bauer,
  B.~M. Murphy, O.~M. Magnussen, E.~E. Fullerton, and O.~G. Shpyrko.
\newblock Photoinduced enhancement of the charge density wave amplitude.
\newblock 117(5):056401.
\newblock Publisher: American Physical Society.

\bibitem{haberer_advances_2002}
Th~Haberer.
\newblock Advances in charged particle therapy.
\newblock In {\em {AIP} Conference Proceedings}, volume 610, pages 157--166.
  American Institute of Physics.
\newblock Issue: 1.

\bibitem{hamm_industrial_2012}
Robert~Wray Hamm and Marianne~Elizabeth Hamm.
\newblock Industrial accelerators and their applications.
\newblock Publisher: World Scientific.

\bibitem{heath_introduction_1995}
David Heath.
\newblock {\em An introduction to experimental design and statistics for
  biology}.
\newblock {CRC} press.

\bibitem{murray_application_2015}
Paul Murray, Fiona Bellany, Laure Benhamou, Dejan-Kresimir Bucar, Alethea
  Tabor, and Tom Sheppard.
\newblock The application of design of experiments ({DoE}) reaction
  optimization and solvent selection in the development of new synthetic
  chemistry.
\newblock 14.

\bibitem{baltz_achievement_2017}
E.~A. Baltz, E.~Trask, M.~Binderbauer, M.~Dikovsky, H.~Gota, R.~Mendoza, J.~C.
  Platt, and P.~F. Riley.
\newblock Achievement of sustained net plasma heating in a fusion experiment
  with the optometrist algorithm.
\newblock 7(1):6425.
\newblock Number: 1 Publisher: Nature Publishing Group.

\bibitem{shahriari_taking_2016}
Bobak Shahriari, Kevin Swersky, Ziyu Wang, Ryan~P. Adams, and Nando de~Freitas.
\newblock Taking the human out of the loop: A review of bayesian optimization.
\newblock 104(1):148--175.

\bibitem{greenhill_bayesian_2020}
Stewart Greenhill, Santu Rana, Sunil Gupta, Pratibha Vellanki, and Svetha
  Venkatesh.
\newblock Bayesian optimization for adaptive experimental design: A review.
\newblock 8:13937--13948.
\newblock Conference Name: {IEEE} Access.

\bibitem{srinivas_gaussian_2010}
Niranjan Srinivas, Andreas Krause, Sham Kakade, and Matthias Seeger.
\newblock Gaussian process optimization in the bandit setting: no regret and
  experimental design.
\newblock In {\em Proceedings of the 27th International Conference on
  International Conference on Machine Learning}, {ICML}'10, pages 1015--1022.
  Omnipress.

\bibitem{duplyakin_active_2016}
Dmitry Duplyakin, Jed Brown, and Robert Ricci.
\newblock Active learning in performance analysis.
\newblock In {\em 2016 {IEEE} International Conference on Cluster Computing
  ({CLUSTER})}, pages 182--191.

\bibitem{smith_less_2018}
Justin~S. Smith, Ben Nebgen, Nicholas Lubbers, Olexandr Isayev, and Adrian~E.
  Roitberg.
\newblock Less is more: Sampling chemical space with active learning.
\newblock 148(24):241733.
\newblock Publisher: American Institute of Physics.

\bibitem{lookman_active_2019}
Turab Lookman, Prasanna~V. Balachandran, Dezhen Xue, and Ruihao Yuan.
\newblock Active learning in materials science with emphasis on adaptive
  sampling using uncertainties for targeted design.
\newblock 5(1):1--17.
\newblock Number: 1 Publisher: Nature Publishing Group.

\bibitem{settles_active_2009}
Burr Settles.
\newblock Active learning literature survey.
\newblock Publisher: University of Wisconsin-Madison Department of Computer
  Sciences.

\bibitem{rasmussen_gaussian_2006}
Carl~Edward Rasmussen and Christopher K.~I. Williams.
\newblock {\em Gaussian processes for machine learning}.
\newblock Adaptive computation and machine learning. {MIT} Press.
\newblock {OCLC}: ocm61285753.

\bibitem{neal_bayesian_2012}
Radford~M Neal.
\newblock {\em Bayesian learning for neural networks}, volume 118.
\newblock Springer Science \& Business Media.

\bibitem{kushner_new_1964}
Harold~J Kushner.
\newblock A new method of locating the maximum point of an arbitrary multipeak
  curve in the presence of noise.

\bibitem{mockus_application_1978}
Jonas Mockus, Vytautas Tiesis, and Antanas Zilinskas.
\newblock The application of bayesian methods for seeking the extremum.
\newblock 2(117):2.

\bibitem{gardner_bayesian_2014}
Jacob~R Gardner, Matt~J Kusner, Zhixiang~Eddie Xu, Kilian~Q Weinberger, and
  John~P Cunningham.
\newblock Bayesian optimization with inequality constraints.
\newblock In {\em {ICML}}, volume 2014, pages 937--945.

\bibitem{conde_research_2017}
Manoel Conde, Sergey Antipov, Darrell Doran, Wei Gai, Qiang Gao, Gwanghui Ha,
  Chunguang Jing, Wanming Liu, Nicole Neveu, John Power, and {others}.
\newblock Research program and recent results at the argonne wakefield
  accelerator facility ({AWA}).
\newblock pages 2885--2887.

\bibitem{huang_review_2007}
Zhirong Huang and Kwang-Je Kim.
\newblock Review of x-ray free-electron laser theory.
\newblock 10(3):034801.
\newblock Publisher: {APS}.

\bibitem{grafstrom_luminosity_2015}
Per Grafström and Witold Kozanecki.
\newblock Luminosity determination at proton colliders.
\newblock 81:97--148.
\newblock Publisher: Elsevier.

\bibitem{hawkes_principles_1996}
Peter~W Hawkes and Erwin Kasper.
\newblock {\em Principles of electron optics}, volume~3.
\newblock Academic press.

\bibitem{zhang_emittance_1996}
Min Zhang.
\newblock Emittance formula for slits and pepper-pot measurement.

\end{thebibliography}

\end{document}